\documentclass[preprint,showpacs,preprintnumbers]{revtex4} 
\usepackage{amsmath} 
\usepackage{amsfonts} 
\usepackage{amssymb}
\usepackage{graphicx} 
\usepackage[usenames]{color} 

\usepackage{bbm}
\def\beq{\begin{equation}}
\def\eeq{\end{equation}}
\def\bea{\begin{eqnarray}}
\def\eea{\end{eqnarray}}
\def\ba{\begin{array}}
\def\ea{\end{array}}

\def\part{\partial}

\begin{document}

\preprint{UdeM-GPP-TH-06-155}
%\preprint{hep-th/0705xxx}

\title{Field theoretic description of the abelian and non-abelian Josephson effect.}
\author{F. Paul Esposito$^1$}
%\email{esposito@physics.uc.edu}  
\author{L.-P. Guay$^2$}
%\email{l-p-guay@umontreal.ca} 
\author{R. B. MacKenzie$^2$} 
%\email{rbmack@lps.umontreal.ca}
\author{M. B. Paranjape$^2$} 
%\email{paranj@lps.umontreal.ca}
\author{L. C. R. Wijewardhana$^1$} 
%\email{rohana@physics.uc.edu}

\affiliation{$^1$Department of Physics,
University of Cinicinnati,
P.O. Box 210011  Cincinnati, OH,
 USA  45221-0011}
\affiliation{$^2$Groupe de physique des particules, D\'epartement de physique,
Universit\'e de Montr\'eal,
C.P. 6128, Succ. Centre-ville, Montr\'eal, 
Qu\'ebec, CANADA H3C 3J7 }

\begin{abstract}  We formulate the Josephson effect in a field theoretic language which affords a straightforward generalization to the non-abelian case.  Our formalism interprets Josephson tunneling as the excitation of pseudo-Goldstone bosons.  We demonstrate the formalism through the consideration of a single junction separating two regions with a purely non-abelian order parameter and a sandwich  of three regions where the central region is in a distinct phase.   Applications to various non-abelian symmetry breaking systems in particle and condensed matter physics are given. 
\end{abstract}
\pacs{11.15.-q, 11.15.Ex }

\maketitle

%%%%%%%%%%%%%%%%%%
{\it Abelian Josephson Effect. }
The abelian Josephson effect \cite{Josephson} concerns  two macroscopic superconductors separated by a thin  layer of normal material, giving rise to a weak interaction between the macroscopic states (condensates of Cooper pairs \cite{Cooper}) of each superconductor. This interaction allows for tunneling-mediated current flow across the junction which is the Josephson effect.

A succinct description of the Josephson effect in terms of effective fields has been given by Feynman \cite{Feynman}. The Cooper pair amplitude of each superconductor should in principle be described by a space- and time-dependent field, but the essence of the effect is captured by using a time-dependent complex amplitude for each superconductor; let these be $\psi(t)$ and $\chi(t)$. Their magnitudes are fixed by the details of the superconductors; for simplicity, we will assume $|\psi|=|\chi|=\sqrt\rho$ below.

If the superconductors are far apart, of course there is no interaction between them (and no Josephson effect). Each state obeys a Schroedinger-like equation:
\begin{eqnarray}
i\hbar\partial_t\psi (t)=E_L\psi(t) \\
i\hbar\partial_t\chi (t)=E_R\chi(t) 
\end{eqnarray}
where $E_L$ and $E_R$ are the chemical potentials on either side.  These equations admit a doubled symmetry, $U(1)\times U(1)$, corresponding to independent phase rotations of the fields $\psi (t)\rightarrow e^{i\theta_\psi}\psi (t)$ and $\chi (t)\rightarrow e^{i\theta_\chi}\chi(t)$.   
As a result, there appear to be two uncoupled massless modes, essentially Nambu-Goldstone (NG)~\cite{Goldstone}  excitations within each superconductor. This enhanced symmetry is in fact an artifact of the treatment of the superconductors as uncoupled systems. As we will see, the introduction of a coupling reduces the symmetry to simultaneous rotation of the fields, $U_D(1)$. This symmetry is spontaneously broken, and the attendant NG boson is absorbed via the Higgs mechanism.

What can be said of the off-diagonal symmetry explicitly broken by the coupling between the superconductors? If the coupling is weak, we still expect a light pseudo-Goldstone mode \cite{pseudo-Goldstone} corresponding to equal and opposite phase rotations of the two fields. In this effective description, this pseudo-Goldstone mode is at the heart of the Josephson effect.

Coupling the superconductors together is described by adding the simplest interaction which preserves the diagonal $U_D(1)$ symmetry.  Since the interaction is weak, the corresponding coupling constant is taken to be a small parameter.  

The simplest coupling of the two superconductors gives the equations:
\begin{eqnarray}
i\hbar\partial_t\psi (t)=E_L\psi(t)+K \chi(t)\label{abelian1}\\
i\hbar\partial_t\chi (t)=E_R\chi(t) +K\psi(t)\label{abelian2}
\end{eqnarray}
This system is  evidently, exactly solvable.  
%Writing $E_L=E+V$ and  $E_R=E-V$ the solution is
%\beq
%\left(\begin{array}{c}\psi(t)\\ \chi(t)\end{array}\right) =
%e^{{-iEt/\hbar}+i\omega t
%\left(\begin{array}{cc}A&B\\ B&-A\end{array}\right)}\left(\begin{array}{c}\psi_0\\ \chi_0\end{array}\right)
%\eeq
The total charge $Q=\psi^*(t)\psi(t)+\chi^*(t)\chi(t)$ is conserved.  However, charge exchange mediated by tunnelling can occur.
Indeed, calculating $Q_\psi =\psi^*(t)\psi(t)$ yields
%\bea\nonumber
%&Q_\psi&=\psi^*(t)\psi(t)=\cos^2\left(\omega t\right)\psi^*_0\psi_0 \\ \nonumber +&\sin^2&\left(\omega t\right)(B\chi^*_0+A\psi^*_0)(B\chi_0+A\psi_0)\\ 
% &+&i\cos\left(\omega t\right)\sin\left(\omega t\right)\left(\left(B\chi^*_0+A\psi^*_0\right)\psi_0 - c.c.\right).
%\eea
\bea\nonumber
Q_\psi &=&\rho\big[\cos^2(\omega t) +\sin^2\left(\omega t\right)\left\{ 1+2AB \cos(\theta_\psi-\theta_\chi)\right\}\\ &~&-\sin (2\omega t)B\sin(\theta_\psi-\theta_\chi)\big].
\eea
where $\psi_0=\sqrt\rho e^{i\theta_\psi}$ and $\chi_0=\sqrt\rho e^{i\theta_\chi}$, $\sqrt\rho $ is the amplitude of the effective field which is the same on both sides (the phases can of course differ), $V=E_L-E_R$ represents a potential applied to the junction, $\omega = {\sqrt{V^2+K^2}/\hbar}$, $A=V/\sqrt{V^2+K^2}$, and $B=K/\sqrt{V^2+K^2}$.  There are two interesting cases to consider.  Firstly with $V=0$ we get the dc Josephson effect
\beq
Q_\psi =\rho\left[1 -\sin (2\omega t)\sin(\theta_\psi-\theta_\chi)\right].
\eeq
The Josephson current is given by
\beq
\dot Q_\psi =\rho\left[2\omega\cos (2\omega t)\sin(\theta_\chi-\theta_\psi)\right] .
\eeq
Writing $\omega =K/\hbar$ and noting that the dc current is valid only for short times, implying $\cos Kt/\hbar\approx 1$, yields
\beq
\dot Q_\psi =\rho\left(2K/\hbar\right) \sin(\theta_\chi-\theta_\psi),
\eeq
the familiar expression for the dc Josephson effect.

Secondly, for the ac effect we take $V\gg K$, which yields
\beq
Q_\psi =\rho\left( 1 -{K\over V}\cos (2V t/\hbar+(\theta_\chi-\theta_\psi))\right)
\eeq
and consequently
\beq
\dot Q_\psi =\rho{2K\over\hbar}\sin (2V t/\hbar+(\theta_\chi-\theta_\psi)).
\eeq
The Josephson acceleration equation follows straightforwardly from the equations of motion for the time dependent phases $\psi(t)=\sqrt\rho e^{i\theta_\psi(t)}$ and $\chi=\sqrt\rho e^{i\theta_\chi(t)}$
\beq
\dot \theta_\chi(t)-\dot\theta_\psi(t) =2V/\hbar .
\eeq
An effective Lagrangian description of the situation is useful for generalization to the non-abelian case.  It is important to realize that it is not really a wavefunction, but a quantum field, that describes each superconductor; these fields are placed into interaction in a Josephson junction.  The effective field theory Lagrangian corresponding to the above analysis is given by
\bea\nonumber
{\cal L} &=& \psi^\dagger i\hbar\dot\psi +\chi^\dagger i\hbar\dot\chi -(\psi^\dagger \chi^\dagger )\left(\begin{array}{cc}{E+V}&0\\ 0&{E-V}\end{array}\right)\left(\begin{array}{c}\psi\\ \chi\end{array}\right) \\ &-&(\psi^\dagger \chi^\dagger )\left(\begin{array}{cc}0&{K}\\ {K}&0\end{array}\right)\left(\begin{array}{c}\psi\\ \chi\end{array}\right) 
\eea
which gives rise to the identical equations of motion as studied before.
In the absence of the coupling term, $K=0$, the symmetry of this model corresponds to independent phase transformations of the two fields $\psi\rightarrow e^{i\zeta}\psi$ and $\chi\rightarrow e^{i\eta}\chi$.  The fact that on either side of the junction the amplitudes of the effective fields are equal and vary negligibly, means that these $U(1)$ symmetries are spontaneously broken, giving rise to what appears to be two NG bosons. As has been mentioned earlier, when $K\ne 0$, only one of these (corresponding to a simultaneous phase rotation of the two fields) is a genuine NG boson and is absorbed via the Higgs mechanism.
What appears to be a second NG boson corresponds to equal and opposite phase rotation of the two fields; the parameter $K$ is the soft breaking parameter of this symmetry, and the corresponding excitations are pseudo-Goldstone bosons.  The frequency associated with these oscillations is correspondingly small, $\omega = {K/ \hbar}$, and describes the usual dc Josephson oscillations.  The ac effect can be seen as an accumulation of the phase $(\theta_\chi-\theta_\psi)\rightarrow 2Vt/\hbar +(\theta_\chi-\theta_\psi)$.

%%%%%%%%%%%%%%%%%%%%%%%%%%%%%%%%%%%%%%%%%%%%%%%%%%%%%%%%%%%%%%%%%%%%%%%%%%%%%%%%
{\it Non-abelian Josephson Effect. }
The non-abelian Josephson effect can now be formulated, in terms of a junction of two systems with spontaneously broken non-abelian gauge symmetries,  which interact with one another very weakly.  Each system should have the same symmetry, or at least the symmetry of one should be contained in the symmetry  of the other.  Hence there is an initial doubling of symmetry.  This doubled symmetry should be spontaneously broken.   The corresponding doubling of the NG bosons that are produced, must be an artefact of the description.  Indeed, coupling the two systems together, so that only the diagonal action of the doubled symmetry generators is preserved, will give rise to one set of NG bosons and one set of pseudo-Goldstone bosons.  
%As the system is considered to be superconducting in a generalized sense, the true NG bosons will be incorporated into the massive vector gauge fields via the Higgs mechanism and will give rise to the corresponding Meissner effect.  
Excitations of the pseudo-Goldstone bosons leads to the non-abelian generalization of the Josephson effect.

It is important to point out that we restrict our attention to Lorentz invariant systems, and especially to Lorentz invariant kinetic terms.   It is only in such systems that the standard counting of Goldstone bosons, which we employ  is valid \cite{smetc}.  Many applications in condensed matter physics and in high density quark/gluon matter in the presence of nonzero chemical potentials do not follow this standard counting.  

The only direct reference to pseudo-Goldstone bosons and the Josephson effect is in a paper of Zhang \cite{Zhang}, 
%where he is considering the $SO(5)$ model 
for the high temperature superconductivity/anti-ferromagnet system \cite{hightc}.  However he does not consider junctions.
%, and the pseudo-Goldstone bosons arising there are a consequence of explicit $SO(5)$ symmetry breaking terms that are added to the effective Lagrangian to push the system away from the $SO(5)$ invariant critical point.  
The Josephson effect in a non-abelian context is also considered in a paper by E. Demler et al. \cite{Berlinsky},
%in the context of $SO(5)$ superconductivity
and in a paper of Ambegoakar et al. \cite{deGennes}, which formulates the problem of a junction for the A phase of liquid $^3$Helium.  However the emphasis in this latter work is not on symmetry considerations but on the geometrical and physical layout that could give rise to such a junction.  

In any case, the way that the abelian or non-abelian Josephson effect corresponds to the excitation of pseudo-Goldstone bosons, has not been explicitly and simply spelled out, which is what we attempt to do here.  We will give two examples which illustrate the non-abelian generalization.  
%%%%%%%%%%%%%%%%%%%%%%%%%%%%%%%%%%%%%%%%%%%%%%%%%%%%%%%%%%%%%%%%%%%%%%%%%%%%%%%%

{\it a) Non-abelian Josephson-like junction with the breaking of $O(N) \rightarrow O(N-1)$}
Consider the junction of two regions where in each region the gauge symmetry is spontaneously broken from $O(N) \rightarrow O(N-1)$.  Coupling the two regions together preserves the diagonal $O(N)_D$ which spontaneously breaks to the diagonal $O(N-1)_D$.  This gives rise to $N-1$ NG bosons, which are eaten by the gauge fields giving them mass.  The explicitly broken off-diagonal symmetries, which would have given rise to $N-1$ other NG bosons,  give rise to $N-1$ pseudo-Goldstone bosons, excitations of which correspond to the non-abelian Josephson effect. 

The Lagrangian we consider contains two real  $N$-component multiplets of scalar fields $\vec\psi$ and $\vec\chi$.  The potential is
\beq
{ V}= {V}^0[\vec\psi]+{ V}^0[\vec\chi] -K(\vec\psi\cdot\vec\chi)
\eeq
where ${ V}^0[\vec\psi]= \lambda (\vec\psi\cdot\vec\psi-a^2)^2$.  
Other coupling terms can in priciple be included, for example $(\vec\psi\cdot\vec\chi)^2$; however the net effect (alignment of order parameters) is the same, so we have not included them in the interest of simplicity.
We do not write in the kinetic terms nor the gauge fields as they play no role in our analysis.  Also, in the effective description of the two regions and their interaction, the spatial dependance is not important.  The minimum of the potential is achieved at the solution of the equations:
\bea\nonumber
4\lambda (\vec\psi\cdot\vec\psi-a^2)\vec\psi -K\vec\chi &=&0\\
4\lambda (\vec\chi\cdot\vec\chi-a^2) \vec\chi -K\vec\psi &=&0
\eea
which has the stable, symmetry breaking solution $\vec\psi =\vec\chi =\sqrt{(K/4\lambda)+a^2}\hat n$ where $\hat n$ is an arbitrary $N$-dimensional unit vector.  Then the second order expansion of the potential evaluated at the minimum is:
\bea\nonumber
{\partial^2 V\over\partial\chi_i\partial\chi_j}={\partial^2 V\over\partial\psi_i\partial\psi_j}&=&K\delta_{ij}+(2K+8\lambda a^2)\hat n_i\hat n_j\\
{\partial^2 V\over\partial\phi_i\partial\chi_j}&=&-K\delta_{ij}
\eea
This $2N\times2N$ matrix  decomposes into $N$ $2\times 2$ matrices, $N-1$ of which correspond to the matrix
\beq
\left(\begin{array}{cc}K&-K\\-K&K\end{array}\right)
\eeq
and
one of which corresponds to
\beq
\left(\begin{array}{cc}3K+8\lambda a^2&-K\\-K&3K+8\lambda a^2\end{array}\right).
\eeq
The eigenspectrum then is 2 heavy modes of energy $8\lambda a^2+2K$ and $8\lambda a^2+4K$, $N-1$ light, pseudo-Goldstone modes of energy $2K$ and $N-1$ zero energy NG modes.  The pseudo-Goldstone modes correspond to equal and opposite non-abelian gauge rotations in the two regions, the NG modes correspond to equal rotations in each region and are removed from the spectrum by the Higgs mechanism and the heavy modes corresponds to oscillations in the lengths of the order parameters. 

{\it b) Josephson sandwich}
In this section we consider a system composed of three regions, each of which has a non-abelian symmetry with partial symmetry breaking. For concreteness, we will consider a three-component real field $\vec\phi_i=(\varphi_i,\psi_i,\eta_i)$ where $i$ denotes the region, though the formalism is more general. The $SO(2)$ subgroup corresponding to the first two components is gauged and is equivalent to ordinary electromagnetism. Spontaneously breaking this subgroup describes superconductivity. The potential is
\bea
V&=&\lambda(|\vec\phi_1|^2-a^2)^2+\lambda(|\vec\phi_2|^2-a^2)^2+
\lambda(|\vec\phi_3|^2-a^2)^2\nonumber \\
&& -g_1(\varphi_1^2+\psi_1^2-\eta_1^2)
+g_2(\varphi_2^2+\psi_2^2-\eta_2^2), \nonumber \\
&& -g_1(\varphi_3^2+\psi_3^2-\eta_3^2)
-K\vec\phi_2\cdot(\vec\phi_1+\vec\phi_3).
\eea
The model has an approximate $SO(3)$ symmetry, which the $g_i$ terms  explicitly, although softly, break  in each region to $SO(2)$. (The sign of the $g_i$ terms is chosen so that positive $g_i$ results in spontaneous breaking of the latter symmetry in regions 1 and 3.) The $K$ terms describe the coupling of adjacent regions, which are assumed to be the true small parameters.  The minimum of the potential  satisfies equations equivalent in form to:
\beq
\left(\begin{array}{ccc}A_\xi&-K&0\\-K&B_\xi&-K\\0&-K&A_\xi\end{array}\right)\left(\begin{array}{c}\xi_1\\ \xi_2\\ \xi_3 \end{array}\right) =0,\label{matrix}
\eeq
where, with the notation $\alpha=4\lambda(|\vec\phi_1|^2-a^2)$ and $\beta = 4\lambda(|\vec\phi_2|^2-a^2)$, we have, for $\xi=\varphi$, $A_\varphi =\alpha -2g_1$ and $B_\varphi =\beta +2g_2$ and for $\xi=\eta$, $A_\eta =\alpha+2g_1$ and $B_\eta =\beta -2g_2$.  A nontrivial solution requires that the determinant of the matrix in equation (\ref{matrix}) vanish for both $\xi=\varphi$ and $\xi=\eta$, which yields the equations
\bea
A_\varphi (A_\varphi B_\varphi -2K^2)&=&0\\
A_\eta (A_\eta B_\eta -2K^2)&=&0.
\eea
The solution for this system of equations is found to be
\beq
\alpha =\sqrt{{2g_1\over g_2}(K^2+2g_1g_2)}, \quad\beta=(g_2/g_1)\alpha .
\eeq
This can be used, in conjunction with the definitions of $\alpha$ and $\beta$ to find the values of $|\vec\phi_1|^2$ and $|\vec\phi_3|^2$ in terms of the coupling constants.  We use the diagonal $SO(3)$ symmetry to rotate the solution into the direction with $\psi_i=0$ without loss of generality.  Then the spectrum of oscillations are governed by the matrix of second derivatives of the potential evaluated at the position of the minimum.  This is a $9\times 9$ matrix which we will only treat perturbatively in $K$, $g_1$ and $g_2$.  The order zero matrix is block diagonal with three, $3\times 3$ blocks, of the form
\beq
8\lambda\left(\begin{array}{ccc}\varphi_i^2&0&\varphi_i\eta_i\\0&0&0\\\varphi_i\eta_i&0&\eta_i^2\end{array}\right)
\eeq
where $i=1,2,3$.  (Note that $\varphi_1=\varphi_3, \eta_1=\eta_3$.)  The eigenvectors of these blocks correspond to six zero modes $v_{a,i},\, a=1,2;i=1,2,3$ and three heavy modes which correspond to radial oscillations and decouple from the low energy theory).  The six normalized light modes can be written using tensor notation as
\beq
\left(\begin{array}{c}0\\1\\0\end{array}\right)\otimes\hat n_i ,\quad \left(\begin{array}{c}\eta_i/\sqrt{\varphi_i^2+\eta_i^2}\\0\\-\varphi_i/\sqrt{\varphi_i^2+\eta_i^2}\end{array}\right)\otimes\hat n_i 
\eeq
where $\hat n_i$ is a unit vector that picks the block with its index $i$, while the three heavy modes are given by
\beq
 \left(\begin{array}{c}\varphi_i/\sqrt{\varphi_i^2+\eta_i^2}\\0\\\eta_i/\sqrt{\varphi_i^2+\eta_i^2}\end{array}\right)\otimes\hat n_i .
 \eeq
To find the spectrum of the light modes we must do degenerate perturbation theory in the six dimensional space defined by the $v_{a,i}$.  This requires us to project the original $9\times 9$ matrix of second derivatives to the degenerate subspace.  This $6\times 6$ matrix decomposes into two $3\times 3$ blocks, identical in form to the matrix of equation (\ref{matrix}), separately for the $v_{1,i}$'s and for the $v_{2,i}$'s.  The matrix for the $v_{1,i}$'s is surprisingly, actually exactly the same as in equation (21) with $\xi=\varphi$, $A_\xi=A_\varphi=\alpha -2g_1$, $B_\xi=B_\varphi =\beta+2g_2$.  For the $v_{2,i}$'s the matrix corresponds to $A_\xi=\alpha$, $B_\xi=\beta$ and $K\rightarrow K^\prime=K (\vec\phi_1\cdot\vec\phi_2/ |\vec\phi_1||\vec\phi_2|)$.

The interesting sector corresponds to the $v_{1,i}$'s since  they involve excitations of the $\psi_i$ degrees of freedom which is required for the transport and production of charge about our choice for the minimum.  The (un-normalized) eigenvectors are
\beq{
v_{1,1}=\left(\ba{c}1\\0\\-1\ea\right) ,\,v_{1,2}=\left(\ba{c}A_\varphi\\-2K\\A_\varphi\ea\right) ,\,v_{1,3}=\left(\ba{c}K\\A_\varphi\\K\ea\right) }
\eeq
with corresponding eigenfrequencies $A_\varphi,A_\varphi+B_\varphi,0$.  $v_{1,1}$ corresponds to the Josephson effect, $v_{1,2}$ corresponds to charge exchange with the intermediate state and $v_{1,3}$ corresponds to the true NG mode which is incorporated into the mass of the photon.  The remaining eigenvectors are
\beq
v_{2,1}=\left(\ba{c}1\\0\\-1\ea\right) ,\,v_{2,2}=\left(\ba{c}\alpha\\-2K^\prime\\ \alpha\ea\right) ,\,v_{2,3}=\left(\ba{c}K^\prime\\ \alpha\\K^\prime\ea\right) 
\eeq
with corresponding eigenfrequencies $\alpha, (\alpha\pm\sqrt {(\alpha -\beta)^2+8(K^\prime)^2})/2$.  These excitations do not correspond to any charge exchange but to oscillations in the amplitude of the superconducting order parameter.  In the limit that $K$ is much smaller than the $g_i$'s, it is clear that $v_{1,1}$ corresponds to the pseudo-Goldstone mode since the corresponding eigenfrequency behaves as $A_\varphi \sim K^2$.  This mode is responsible for the tunnelling of charge across region 2 since it corresponds to perturbing $\psi_1$ in an equal and opposite direction to $\psi_3$.  The charge operator is $Q_i\sim i(\varphi_i{d\over dt}\psi_i -\psi_i{d\over dt}\varphi_i)\sim i\varphi_i{d\over dt}\delta\psi_i$ since $\psi_i$ vanishes to zero order.

{\it Conclusions. }
We have formulated the Josephson effect in the language of effective Lagrangians.   This allowed for a generalization to non-abelian symmetries and the corresponding non-abelian Josephson effect.   We find that the Josephson effect corresponds to the excitations of pseudo-Goldstone bosons.    First we applied our formalism to the breaking of an $SO(N)$ symmetry resulting in the Josephson effect in $N-1$ non-abelian charges.  Then we considered a sandwich of three regions with an underlying, approximate  $SO(3)$ symmetry, that is explicitly broken to $SO(2)$.   The unbroken $SO(2)$ is gauged and the explicit $SO(3)$ breaking terms drive the end regions to spontaneously break the $SO(2)$ symmetry making them superconducting. In the intermediate region the explicit $SO(3)$ breaking terms do not cause spontaneous breaking of the $SO(2)$.   We find that the intermediate region mediates the exchange of charge between the two end regions, giving rise to a Josephson effect.  

Our formalism could be applied to physical situations involving the spontaneous breaking of non-abelian gauged symmetries, or even co-existing abelian symmetries, if an interface arises.   The interplay between multiple order parameters at the effective lagrangian level for strongly interacting theories has already been used to resolve basic puzzles  between confinement and chiral symmetry \cite{mst}.   Indeed, in other symmetry breaking scenarios, it can happen that
dimension four interactions, which we have not considered for simplicity, can become 
relevant, see \cite{mst} for details.  One promising area where this could occur is in the high density phases of QCD (for a review see \cite{Rajagopal}) which are expected to occur in contiguous regions of neutron stars.   Other examples of situations wherein non-minimal order parameters may provide a venue for our formalism include two-band superconductors \cite{bouquet}, $d$-wave high $T_C$ superconductors \cite{hightc},  $p$-wave heavy-fermion superconductors \cite{Steglich}, the A phase of liquid $^3$Helium \cite{deGennes,Volovik}, and nonlinear optics, where complicated order parameters often occur \cite{kivshar}.  

{\it  Acknowledgments }
We thank NSERC of Canada and USDOE Grant DE-FG02-84ER40153 for financial support.  

%%%%%%%%%%%%%%%%%%%%  BIBLIO  %%%%%%%%%%%%%%%%%%%%%%%%%%%

\end{document}